# Competition between superconductivity and spin density wave

Tian De Cao

Department of Physics, Nanjing University of Information Science & Technology, Nanjing 210044, China

**Abstract**

The Hubbard model has been investigated widely by many authors, while this work may be new in two aspects. One, we focus on the possible effects of the positions of the gaps associated with the pairing and the spin density wave. Two, we suggest that the models with different parameters are appropriate for different materials (or a material in different doped regions). This will lead to some new insights into the high temperature superconductors. It is shown that the SDW can appear at some temperature region when the on-site Coulomb interaction is larger, while the SC requires a decreased U at a lower temperature. This can qualitatively explain the relationship between superconducting and pseudogap states of Cu-based superconductors in underdoped and optimally doped regions. The superinsulator is also discussed.

## 1. Introduction

Many scenarios for the pseudogap have been suggested, such as the pairs-based scenario [1,2,3] and the spin ordering scenario [4,5]. Some experiments showed that two orders may coexist in some superconductors [6,7]. However, the pseudogap remains controversial and requires further studies. We suggested that some gap has occupied the Fermi level in the antinodal region of cuprate superconductors before the pairing gap appears in the pseudogap state ($\Delta_{pair}(\vec{k}) \neq 0$ is for $|\vec{k} - \vec{k}_F| >> 0$) [8]. **The Fermi level (chemical potential position) is determined by $E(k, \Delta_k) = 0$ for $\Delta_k = 0$, $\Delta_k$ is either the pairing gap or other gap. The wave vectors are called as the Fermi wave vector in this case, $k \to \vec{k}_F$. While we find $E(k, \Delta_k) \neq 0$ for $\Delta_k \neq 0$, thus the** **quasiparticles could not be defined around the Fermi level for $\Delta_k \neq 0$, the Fermi surface is broken. In this case, we say that the gap occupies the Fermi level.** We think that the position of a gap is the key to understand the high temperature superconductivity, and how and why the pairing gap varies in a material has to be investigated. This investigation will reveal that there should be a crossover from the SDW dominated pseudogap to the SC state in Cu-based superconductors when they develop from the underdoped region to the optimally doped region or from the high temperature range to the low temperature range, and the possible existence of a critical point [9] is also supported by this investigation.

**2. Calculations and results**

Let us first discuss the model parameters. Although one expects that the properties of high temperature superconductors may be included in a model with the fixed parameters, the electronic structures of these materials must be changed with doping, and then the model parameters should be changed with doping. In the Hubbard model to describe the physics of materials, the hopping matrix element $t$ will decrease with the increase of $U$ when the electrons (holes) tend toward being localized, and this can be found in the derivations of a Hamiltonian. Thus **a large $U$ model (or the t-j model) should be used to describe antiferromagnetic insulator, and we will not discuss this case.** When we consider a metal, $U$ should be not too large (although the non-half-filled large-U model includes the metal), in this case, other interactions should be considered also. However, we will qualitatively model some high temperature superconductors, the Hubbard model (U>0) is still considered, other interactions are neglected, and the Hubbard model can be written in

$$H = \sum_{k,\sigma} \xi_k d^+_{k\sigma} d_{k\sigma} + \sum_q U\hat{\rho}(q)\hat{\rho}(-q) - \sum_q U\hat{S}_z(q)\hat{S}_z(-q) \tag{1}$$

where $\hat{\rho}(q) = \frac{1}{2}\sum_{k,\sigma} d^+_{k+q\sigma} d_{k\sigma}$ and $\hat{S}(q) = \frac{1}{2}\sum_{k,\sigma} \sigma d^+_{k+q\sigma} d_{k\sigma}$ are introduced. We have neglected the vector symbol of the wave vectors. In some detail problems, other models may be used, but we find that the pseudogap can be qualitatively described with the Hubbard model. The Hubbard model is widely used in condensed matter physics, but our calculation has some different aspects, particularly, our focus is on the positions of the energy gaps. To discuss the SDW states with the dynamic equation method, we define these functions

$$G(k\sigma, \tau - \tau') = - <T_\tau d_{k\sigma}(\tau) d^+_{k\sigma}(\tau')> \tag{2}$$

$$F^+(k\sigma, \tau - \tau') = <T_\tau d^+_{k\sigma}(\tau) d^+_{\bar{k}\bar{\sigma}}(\tau')> \tag{3}$$

$$F(k\sigma, \tau - \tau') = <T_\tau d_{\bar{k}\bar{\sigma}}(\tau) d_{k\sigma}(\tau')> \tag{4}$$

$$D(k, Q, \sigma, \tau - \tau') = <T_\tau d^+_{k\sigma}(\tau) d_{k-Q\bar{\sigma}}(\tau')> \tag{5}$$

The function $D$ describes the transverse SDW states. The longitudinal SDW states will not be discussed because they can be described equivalently in terms of the longitudinal spin polarization. Using the dynamic equation method [8], after a lengthy calculation to the second order of approximation, we arrive at these three equations

$$[-i\omega_n + \xi_k + \Xi^{(-)}(k, \sigma, i\omega_n)] G(k\sigma, i\omega_n)$$
$$= -\lambda(k, -i\omega_n) - \Delta^{(+)}_{pair}(k, \sigma, i\omega_n) F^+(\bar{k}\bar{\sigma}, i\omega_n) + \Delta^{(+)}_{SDW}(k, Q, \sigma, i\omega_n) D(k, Q, \sigma, -i\omega_n) \tag{6}$$

$$[-i\omega_n - \xi_k + \Xi^{(+)}(k, \sigma, i\omega_n)] D(k, Q, \sigma, i\omega_n)] = \Delta^{(-)}_{SDW}(k, \sigma, i\omega_n) G(k - Q, \sigma, -i\omega_n) \tag{7}$$

$$[-i\omega_n - \xi_k + \Xi^{(+)}(k, \sigma, i\omega_n)] F^+(k\sigma, i\omega_n) = -\Delta^{+(-)}_{pair}(k, \sigma, i\omega_n) G(\bar{k}\bar{\sigma}, i\omega_n) \tag{8}$$

where $\lambda(k, \omega) = [1 - Un/2(\omega + \xi_k)]$, $\Xi^{(\pm)}(k, \sigma, i\omega_n) = \sum_q P(k, q, \sigma)/(i\omega_n \pm \xi_{k+q})$, and

$$P(k, q, \sigma) = U^2 <\hat{S}(-q)\hat{S}(q)> - 2\sigma U^2 <\hat{\rho}(-q)\hat{S}(q)> + U^2 <\hat{\rho}(-q)\hat{\rho}(q)> \tag{9}$$

$$\Delta^{(\pm)}_{pair}(k, \sigma, i\omega_n) = U \sum_q \frac{\xi_{k+q} - \xi_k}{-i\omega_n \pm \xi_{k+q}} F(k+q\sigma, \tau=0) \tag{10}$$

$$\Delta^{+(\pm)}_{pair}(k, \sigma, i\omega_n) = U \sum_q \frac{\xi_{k+q} - \xi_k}{-i\omega_n \pm \xi_{k+q}} F^+(k+q\sigma, \tau=0) \tag{11}$$

$$\Delta^{(\pm)}_{SDW}(k, Q, \sigma, i\omega_n) = U \sum_q \frac{\xi_{k+q-Q} - \xi_{k-Q}}{-i\omega_n \pm \xi_{k+q}} D(k+q, Q, \sigma, \tau=0) \tag{12}$$

where $n$ is the electron (quasi-particle) number at each site. Some constant numbers have been absorbed into the chemical potential. To obtain Eq. (7), we have neglected one term including $\xi_{k+Q} - \xi_k \approx 0$ (as supported below) near the Fermi level. We will find that the wave vector $Q$ is approximately $(\pi,\pi)$. Using $\varepsilon(k+Q) = \varepsilon(k)$ (this will be confirmed below) for the excitation energies in the SDW state, we have $D(k+q, -Q, \sigma, 0) =$ $D(k+q,Q,\sigma,0)$ and $G(k-Q,\sigma,i\omega_n) = G(k,\sigma,i\omega_n)$. It seems appropriate for us to take $F^+ = F$, thus $\Delta^+ = \Delta$. The correlation function $P(k,q,\sigma)$ could be calculated by the same method if we were interested in the quantitative results, but it is not difficult to find how $P$ depends qualitatively on the model parameters by its definition. The contributions of the energies near (and on) the Fermi level are considered below. For simplicity, we can discuss the related results by letting either $\Delta_{pair} \to 0$ or $\Delta_{SDW} \to 0$ respectively, and then the co-existence of the pairs and the SDW state can be also found in these results. We will not consider the possible ferromagnetism, thus the $\sigma$ dependence of functions will be neglected.

Taking $\Delta_{pair}^{(+)} = 0$, we obtain the SDW gap function equation

$$\Delta_{SDW}^{(-)}(k,Q,\omega)$$

$$= U\sum_q \frac{\xi_{k+q-Q} - \xi_{k-Q}}{-\omega - \xi_{k+q}} \cdot [\frac{\Delta_{SDW}^{(-)}(k+q,Q,\varepsilon_{k+q}^{(+)})\lambda(k+q,\varepsilon_{k+q}^{(+)})n_F(\varepsilon_{k+q}^{(+)})}{\omega^{(+)}(\varepsilon_{k+q}^{(+)}) - \omega^{(-)}(\varepsilon_{k+q}^{(+)})}$$

$$- \frac{\Delta_{SDW}^{(-)}(k+q,Q,\varepsilon_{k+q}^{(-)})\lambda(k+q,\varepsilon_{k+q}^{(-)})n_F(\varepsilon_{k+q}^{(-)})}{\omega^{(+)}(\varepsilon_{k+q}^{(-)}) - \omega^{(-)}(\varepsilon_{k+q}^{(-)})}] \qquad (13)$$

where $\varepsilon_k^{(\pm)}$ are determined by the equations $\varepsilon_k^{(\pm)} = \omega^{(\pm)}(k,\varepsilon_k^{(\pm)})$ and

$$\omega^{(\pm)}(k,\varepsilon_k) = -\xi_k + \Xi^{(+)}(k,\sigma,\varepsilon_k) \pm |\Delta_{SDW}^{(-)}(k,Q,\sigma,\varepsilon_k)| \qquad (14)$$

The features of the SDW state could be found in Eqs.(13) and (14), but these results can be shown in a more obvious equation. The SDW temperature $T_{SDW}^*$ is determined by letting $\Delta_{SDW} \to 0$ in the right of Eq.(13), while the wave vector $Q$ is determined by finding the highest $T_{SDW}^*$. Let $\chi_k$ meet the equation

$$\chi_k = -\xi_k + \Xi^{(+)}(k,\chi_k) \qquad (15)$$

and provided $\Delta_{SDW}^{(-)}$ has the non s-wave symmetry which will be confirmed below, we simplify Eq.(13) to the approximate form

$$\Delta_{SDW}^{(-)}(k,Q,\chi_k) = U\sum_q \frac{\xi_{k+q-Q} - \xi_{k-Q}}{-\chi_k - \xi_{k+q}} \Delta_{SDW}^{(-)}(k+q,Q,\chi_k) \frac{\partial}{\partial \chi}[\lambda(k+q,\chi)n_F(\chi)]\Big|_{\chi=\chi_k} \qquad (16)$$

Having noted $\lambda(k,\chi) = [1 - Un/2(\chi + \xi_k)]$, we understand why $\Delta_{SDW}^{(-)}$ favors appearing on the Fermi level with Eq.(16), thus $T^*_{SDW}$ is determined by

$$\Delta^{(-)}_{SDW}(k_F,Q,0)=U\sum_q \frac{\xi_{k_F-Q}-\xi_{k_F+q-Q}}{\xi_{k_F+q}} \Delta^{(-)}_{SDW}(k_F+q,Q,0)\frac{\partial}{\partial \chi}[\lambda(k_F+q,\chi)n_F(\chi)]\big|_{\chi=0} \quad (17)$$

Some results can be easily found on this equation. We estimate that the highest $T^*_{SDW}$ requires the characteristic wave vector $Q=(\pi, \pi)$ with Eq. (17), some reason is also found soon, but we first use $Q=(\pi, \pi)$ to discuss some problems. The non s-wave symmetry of gaps (similar to the d-wave symmetry) can be found in the square lattice model. For example, the nearest-neighbors hopping on a square lattice leads to the dispersion relation $\xi_k = -2t(\cos k_x + \cos k_y) - \mu$. We have $\mu=0$ for the half-filled band at $U=0$. As an evaluation, we can take $\mu=0$ for the model near the half-filled band at $U>0$ [10], thus $\xi_k=0$ gives $\vec{k} = (\pi/2, \pi/2)$ or $(\pi, 0)$ at which the Eq.(17) is rewritten in

$$\Delta^{(-)}_{SDW}(k_F,Q,0)=U\sum_q \Delta^{(-)}_{SDW}(k_F+q,Q,0)\frac{\partial}{\partial \chi}[\lambda(k_F+q,\chi)n_F(\chi)]\big|_{\chi=0} \quad (18)$$

Because $\xi_{k+q}=2t(\sin q_x + \sin q_y)$ at $k=(\pi/2, \pi/2)$ and $\xi_{k+q}=2t(\cos q_x - \cos q_y)$ at $k=(\pi, 0)$. The wave vectors $q \in BZ$, and the summary of the right of Eq. (18) for $\vec{k}=(\pi/2, \pi/2)$ is similar to the integral where the integrand is an odd function, and the integral is performed over an interval symmetric with respect to the origin point, thus $\Delta^{(-)}_{SDW}=0$ for $\vec{k}=(\pi/2, \pi/2)$, and we conclude $\Delta^{(-)}_{SDW}$ favors appearing in the antinodal region around $\vec{k}=(\pi, 0)$. Moreover, as used above, it is easy to find $\xi_{k+Q}-\xi_k \approx 0$ in the antinodal region, if this is used, Eq. (14) shows $\varepsilon_{k+Q}-\varepsilon_k \approx 0$, which is just what the SDW state requires. The SDW state of Eq. (17) usually requires $\frac{\partial}{\partial \chi}[\lambda(k_F+q,\chi)n_F(\chi)]\big|_{\chi=0} >0$, while this requires that the temperature is not too low for the SDW state, the SDW state may disappear when the temperature is low enough. Therefore, the SDW gap initially increases and afterwards decreases with decreasing temperature. The solutions above will become corrupted when the on-site interaction $U$ is too large (no Fermi surface at the half-filled model), but Eq. (17) shows that the SDW temperature increases with the increasing of $U$. Therefore, we arrive at these conclusions:

(1) The Hubbard model includes the spin density waves.

(2) The SDW gap favors appearing around the chemical position.

(3) The SDW intends to disappear at low enough temperature for some model parameters.

(4) The SDW gap favors appearing in the antinodal region for the square lattice.

(5) The SDW temperature intends to increase with $U$.

These conclusions (1)-(5) should be affected by the model parameters and the band filling, and the conclusions (2), (3) and (5) are important in understanding the competition between the high temperature superconductivity and the SDW state.

In addition, to arrive at the high $T_{SDW}^*$, Eq.(13) shows that $\Delta_{SDW}(\vec{k})$ should be zero on the most segments of the Fermi surface (strictly speaking, which consists of the chemical potential positions because the Fermi surface is broken by energy gaps), thus the non-s wave SDW gaps (they may have the d-wave symmetry) favor the high SDW temperature. This shows that other possible s-wave SDW gap corresponds to the very low $T_{SDW}^*$. The electronic structures of underdoped cuprates are highly anisotropic, and the SDW temperatures are very high, this may be explained with the results above (larger $U$ and higher $T_{SDW}^*$). The conclusion (3) shows that the SDW gap initially increases and then decreases with the decrease of temperature; this will be considered for the crossover from the pseudogap state to the superconducting state of high temperature superconductors.

Now we let $\Delta_{SDW}=0$, this leads to the pair function equation

$$\Delta_{pair}^{(\pm)}(k,\omega)$$

$$=U\sum_{q}\frac{\xi_{k+q}-\xi_k}{-\omega\pm\xi_{k+q}}\cdot[\frac{\Delta_{pair}^{(-)}(k+q,E_{k+q}^{(+)})\lambda(k+q,-E_{k+q}^{(+)})n_F(E_{k+q}^{(+)})}{\Omega^{(+)}(E_{k+q}^{(+)})-\Omega^{(-)}(E_{k+q}^{(+)})}$$

$$-\frac{\Delta_{pair}^{(-)}(k+q,E_{k+q}^{(-)})\lambda(k+q,-E_{k+q}^{(-)})n_F(E_{k+q}^{(-)})}{\Omega^{(+)}(E_{k+q}^{(-)})-\Omega^{(-)}(E_{k+q}^{(-)})}] \qquad (19)$$

where $E_k^{(\pm)}=\Omega^{(\pm)}(k,E_k^{(\pm)})$ and

$$\Omega^{(\pm)}(k,\omega)$$

$$=\frac{1}{2}\{\Xi^{(-)}(k,\omega)+\Xi^{(+)}(k,\omega)\pm\sqrt{[2\xi_k+\Xi^{(-)}(k\omega)-\Xi^{(+)}(k\omega)]^2+4\Delta_{pair}^{(-)}(k\omega)\Delta_{pair}^{(+)}(k\omega)}\} \qquad (20)$$

Equation (20) shows that the pairing gap function is determined by $g_{pair}(k) = 2\sqrt{\Delta_{pair}^{(-)}(k,E_k)\Delta_{pair}^{(+)}(k,E_k)}$ which is more complex than the well-known one. With the discussion similar to the SDW state, we obtain these results:

1) The Hubbard model includes the singlet pairing state.

2) The pairing gap favors appearing around the chemical potential position.

3) The pairing gap favors appearing in the antinodal region for the square lattice.

4) The highest pairing temperature requires an appropriate $U$.

Our conclusions are limited to the half-filled band: they must be changed with the band filling (or doping in materials). However, we will not directly discuss the effect of doping, we focus on the changes of model parameters with the doping, and we discuss how the features of superconductors change with the model parameters. It is necessary to note that the on-site interaction decreases with the increased doping in cuprates when we model the changes of the properties of the cuprates with the Hubbard model.

Two different conditions of pairs and SDW are the most important results. One, when $U$ is so large that $E_k^{(+)} - E_k^{(-)} \neq 0$ for $g_{pair}(k) = 0$ for any wave vectors, the pairing temperature $T_{pair} \to 0K$. That is to say, the highest $T_{pair}$ requires a moderate parameter $U$ (and $t$). In contrast with the pairing state, the SDW temperature intends to increase with increasing $U$ (but it is not too large, otherwise, no Fermi surface in the half filling case.). Two, the pairing state favors the decreasing temperature, while the SDW intends to disappear at low enough temperature. This is very interesting because it is benefit to the SC state.

In the above, we have considered the SDW state and the pairing state, respectively. If we considered the two kinds of states at the same time, on the basis of these conclusions above, it is easy to find that the pairing state intends to overcome the SDW state when the temperature is low enough, and this is in agreement with the results observed in many superconductors as we will discuss below. On the basis of these results, it is deduced that the SDW gap should first appears around the Fermi surface when the on-site interaction is large enough, while the pairing gap may appear near the SDW gap (it will be weakened with decreasing U) after the temperature is further decreased. There should be a crossover parameter region from SDW to SC. In this crossover region, if the paring gap occupies some part of the Fermi surface, the coexistence between SC and SDW appear. If the pairing gap is near to the SDW gap which around the Fermi surface, the pairing gap is far away from the Fermi surface, and thus the preformed pairs (PP) appear, and this is the case of the coexistence between PP and SDW. This shows $T_{SDW}^* > T_c$. Following the decrease of the on-site interaction, $T_c \geq T_{SDW}^*$ may be obtained. Therefore, the SDW may hinder the superconductivity from occurring in the room temperature under the usual conditions.

## 3. Explanations of experiment results

If these results are used for discussing cuprate superconductors, there should be the crossover from the SDW dominated pseudogap to the pairs dominated gap when a cuprate superconductor undergoes the transition from the underdoped region to the optimally doped region or from high temperature to low temperature. This allows the possible existence of a critical point because the SDW temperature decreases with the increasing doping of high temperature superconductors. There may be the pure SDW phase, the coexisting phase of the SDW and the pairs, and the pure pairing phase in some cuprate superconductors. These could be deduced on the basis of the calculations above. This deduction is consistent with the results observed in experiments.

To include other superconductors, like iron-based superconductors, we should make some deductions. We divide the pseudogap $g$ into two parts, $g = g_{pair} \oplus g_{other}$, but $g_{other} = g_{SDW}$ or $g_{band}$ or both of them. $g_{pair}$ expresses the pairing gap while $g_{SDW}$ is the SDW gap. $g_{pair}$ will appear near the position of $g_{other}$ in the momentum space, $g_{other}$ could be small but be around the Fermi level. $g_{pair}(\vec{k}) \neq 0$ is for $\vec{k} \in$ pairing gap space (PGS) while $g_{SDW}(\vec{k}) \neq 0$ is for $\vec{k} \in$ SDW gap space (SGS), because the SGS occupy around the Fermi level at $T > T_{pair}$, thus the PGS is not over the Fermi level in the pseudogap state. In cuprates, $g$ should appear in the antinodal region for underdoped cuprates. $g_{SDW}$ appears at the temperature $T \leq T_{SDW}^*$ while $g_{pair}$ appears at the temperature $T < T_{pair}^* < T_{SDW}^*$ for the underdoped region in which the non-s symmetry of $g_{SDW}(\vec{k})$ is very obvious. However, $g_{SDW}$ and its distribution region (SGS) should intend to be weakened when a superconductor develops from the underdoped region to the optimally doped region following the increase of the s-wave part of energy gaps.

Let us now discuss some experiments.

Ex.1: One of the interesting ARPES experiment results is the one reported by Kondo and coauthors [11]: they observed the almost perfect linear anti-correlation between the coherent spectral weight $W_{CP}$ and the pseudogap opening $W_{PG}$, and they gave the conclusion "competition between the pseudogap and superconductivity in the high-$T_c$ copper oxides". However, we find that this experiment can be explained with the so-called "two-gap mechanism". On the basis of the discussion above, with the increasing doping or the decreasing temperature, the PGS should shift toward the Fermi level (antinodal region first, nodal region afterward). Therefore, the coherent spectral weight $W_{CP}$ increases as the PGS shifts toward the Fermi level ($W_{PG}$ decreases at the same time). Thus, we can understand the almost perfect linear anti-correlation between $W_{CP}$ and $W_{PG}$. That is to say, the Kondo's experiment may confirm the two-gap mechanism for the pseudogap.

Ex.2: In the past, one suggested that superconducting pairing fluctuations be responsible for the pseudogap. Martin and Balatsky proposed a probe of pseudogap by Josephson tunneling [12], Bergeal and coauthors tested this proposition [13] and conclude that superconducting pairing fluctuations could not explain the opening of the pseudogap at higher temperature. However, as we discussed in our calculations and Ex.1, the pairs based gap should appear in $T < T_{pair}^* < T_{SDW}^*$ for a underdoped region, the pairs based gap with the non-s wave symmetry should be far away from the Fermi level, the Josephson current should be found at low enough temperature when the pairing gap appears. With increasing doping, the Josephson current should be found at high and high temperature because the pairs based gap shifts toward the Fermi level, and this is observed by Bergeal's experiment. Of course, the pairing gap far away from the Fermi level in my view is different from the pairing fluctuations, because whether the latter could appear over the Fermi level has not been argued in theory, while we suggest that the pairs based gap far away from the Fermi level also contributes to the pseudogap. To see a large Josephson current in pseudogap state, one should increase the DC voltage. That is to say, we believe that Bergeal and coauthors confirmed the two-gap mechanism of the pseudogap.

Ex.3: Some neutron diffraction experiments or the zero-field muon spin relaxation measurements observed the magnetic order or the magnetic excitation in the pseudogap phase of high-$T_C$ superconductors [14,15,16]. However, as we discussed above, $g_{other}$ is around the Fermi level and the SGS could be due to magnetic orders. That is to say, the preformed pairs could coexist with a weak magnetic order on the basis of our calculations.

Other examples: two orders may be observed in some superconductors, and this is allowed in the theory as discussed above. Moreover, the supperinsulator is discovered in experiment [17] which shows that the resistivity is infinity below a transition temperature. According to our theory, if $g_{other}$ is around the whole Fermi surface, for which the gap has the s-wave symmetry, the material is an insulator. If a s-wave pairing gap appears aside the gap, and $g_{pair}$ is far away from the Fermi surface, the material is not in a superconducting state. Because $g_{pair} + g_{other}$ (or $g_{SDW} + g_{band}$) provide a large gap, the supperinsulator can be understood theoretically and observed in experiment. As discussed above, the s-wave $g_{pair}$ (or $g_{SDW}$) must appear at low temperature, this is in agreement with the experiment, and this also explains why supperinsulator appears at very low temperature.